\documentclass[prl,twocolumn,showpacs,preprintnumbers,amsmath,amssymb,superscriptaddress]{revtex4-1}
\usepackage{graphicx}
\usepackage{dcolumn}
\usepackage{bm}
\usepackage{color}

\begin{document}

\title{
Small-q Phonon Mediated Unconventional Superconductivity in the Iron Pnictides}

\author{A. Aperis}\email{aaperis@central.ntua.gr}
\affiliation{Department of Physics, National Technical University of Athens, GR-15780 Athens, Greece}
\author{P. Kotetes}
\affiliation{Department of Physics, National Technical University of Athens, GR-15780 Athens, Greece}
\affiliation{Institut f\"{u}r Theoretische Festk\"{o}rperphysik, Karlsruhe Institute of Technology, 76128 Karlsruhe, Germany}
\author{G. Varelogiannis}
\affiliation{Department of Physics, National Technical University of Athens, GR-15780 Athens, Greece}
\author{P. M. Oppeneer}
\affiliation{Department of Physics and Materials Science, Uppsala University, Box 530, S-751 21 Uppsala, Sweden}
\vskip 0.7cm


\begin{abstract}
We report self-consistent calculations of the gap symmetry for the iron-based high-temperature superconductors using realistic small-q phonon mediated pairing potentials and four-band energy dispersions. When both electron and hole Fermi surface pockets are present,
we obtain the nodeless $s_\pm$ state that was first encountered
in a spin-fluctuations mechanism picture.
Nodal $s_\pm$ as well as other gap structures such as $d_{x^2-y^2}$, $s_\pm+d_{x^2-y^2}$ and even a $p$-wave {\it triplet}
state, are accessible upon doping within our phononic mechanism.
Our results {\it resolve} the conflict between phase sensitive experiments reporting
a gap changing sign attributed previously only to a non-phononic mechanism and isotope effect measurements proving the involvement of phonons in the pairing.

\end{abstract}

\maketitle

One of the foremost issues in contemporary condensed-matter physics is the nature of the medium-high-temperature superconductivity (SC) in the recently discovered iron pnictides \cite{Kamihara}.
The focal point of current research is to understand whether the bosonic pairing that drives the SC is due to spin-fluctuations or phonons, and how the pairing mechanism concurs with the emerging symmetry of the superconducting gap \cite{Mazin2}.
Experimental techniques have recently given controversial results \cite{NatureNeutrons,IsotopeNat}.
Angular-resolved photoemission spectroscopy and Andreev spectroscopy indicate one nearly isotropic gap or two isotropic gaps
\cite{BCS-likeSmFeAsF,Ding,Kondo}, whereas penetration depth \cite{Martin,Malone} and nuclear magnetic resonance (NMR) \cite{Grafe,Yashima} measurements support both nodeless and nodal gap structures. Moreover, very recent experiments on the evolution of the SC gap symmetry of LaFeAsO$_{1-x}$F$_x$ with doping, conclude that there is a transition from nodeless to nodal SC upon moving from the underdoped to the overdoped regime of this compound \cite{YukeLi}. Notably, a nodeless gap is not necessarily a conventional isotropic $s$-wave gap here.

An important ingredient for discussing the gap symmetry is the Fermi surface (FS) of both 1111 (LaOFeAs) and 122 (BaFe$_2$As$_2$) parent compounds,
which is remarkably simple. The strongly two-dimensional FS mainly consists of two electron and two hole pockets, with a nesting wavevector $\bf{Q}=(\pi,\pi)$ connecting them in the folded Brillouin zone (BZ) \cite{Singh1,Mazin1} 
 Fermi surface nesting stabilizes a spin density wave which yields to superconductivity upon doping \cite{Clarina}. If
spin-fluctuations mediate the pairing then a nodeless unconventional superconducting gap
reversing its sign between the electron and hole FS sheets is expected, known as the $s_\pm$ state
\cite{Mazin1,Kuroki,Graser,Mazin2}.

Phase sensitive experiments are needed to distinguish the $s_\pm$
state from the conventional isotropic $s$-wave state
and indeed, inelastic neutron scattering results appear to be compatible with the coherence factors of the $s_\pm$ state \cite{NatureNeutrons}. This has been interpreted as evidence for a
non-phononic pairing mechanism probably involving spin-fluctuations.
However, a systematic NMR study of the dependence of low-lying excitations upon fluorine doping in LaFeAsO$_{1-x}$F$_x$ has shown that antiferromagnetic (AFM) fluctuations are suppressed with doping while $T_c$ remains unaffected, implying that AFM fluctuations may be irrelevant to SC \cite{Nakai}.
Most importantly, isotope effect measurements \cite{IsotopeNat} directly prove
the involvement of the phonons in the pairing mechanism. Additional experimental
findings do support the relevance of phonons in the pairing \cite{Granath}.
However, it is mysterious how this could be compatible
with a gap that changes sign.

In this Letter we propose a new picture for superconductivity in the iron pnictides on the basis of self-consistent calculations of the gap symmetry.
We argue, first, that
small-q electron-phonon scattering dominates the pairing.
Second, we show that this solely phononic mechanism in the presence of strong Coulombic interactions can naturally produce the unconventional gap symmetries proposed for the
pnictides.
When both electron and hole FS pockets are present,
the nodeless $s_\pm$ gap is obtained self-consistently, because of the intricate four band pnictide FS combined with strong interband scattering.
Third, nodal gap structures may also develop when going from underdoped to overdoped regimes upon doping. These include $s_\pm$, $d_{x^2-y^2}$-wave, $s_\pm+d_{x^2-y^2}$ and even $p$-wave triplet SC.

The effective electron-phonon interaction may be dominated by the forward processes only when the
Coulomb interaction is short range or Hubbard-like \cite{smq1,smqGV1}, which is a requirement for  antiferromagnetism as well. Therefore, the proximity of the AFM
and small-q phonon mediated SC phases is
natural, and this is indeed the case in the pnictides.
Unconventional SC due to small-q phonon pairing has already been considered
for high-T$_c$ cuprates 
\cite{Abrikosov,smqGV1,smqGV2}, heavy fermion
\cite{Agterberg}, organic \cite{smqGV3} and cobaltite \cite{Cobaltite} systems.

\begin{figure}[t!]
\centering
\begin{minipage}[b]{1.5in}
\includegraphics[width=\textwidth]{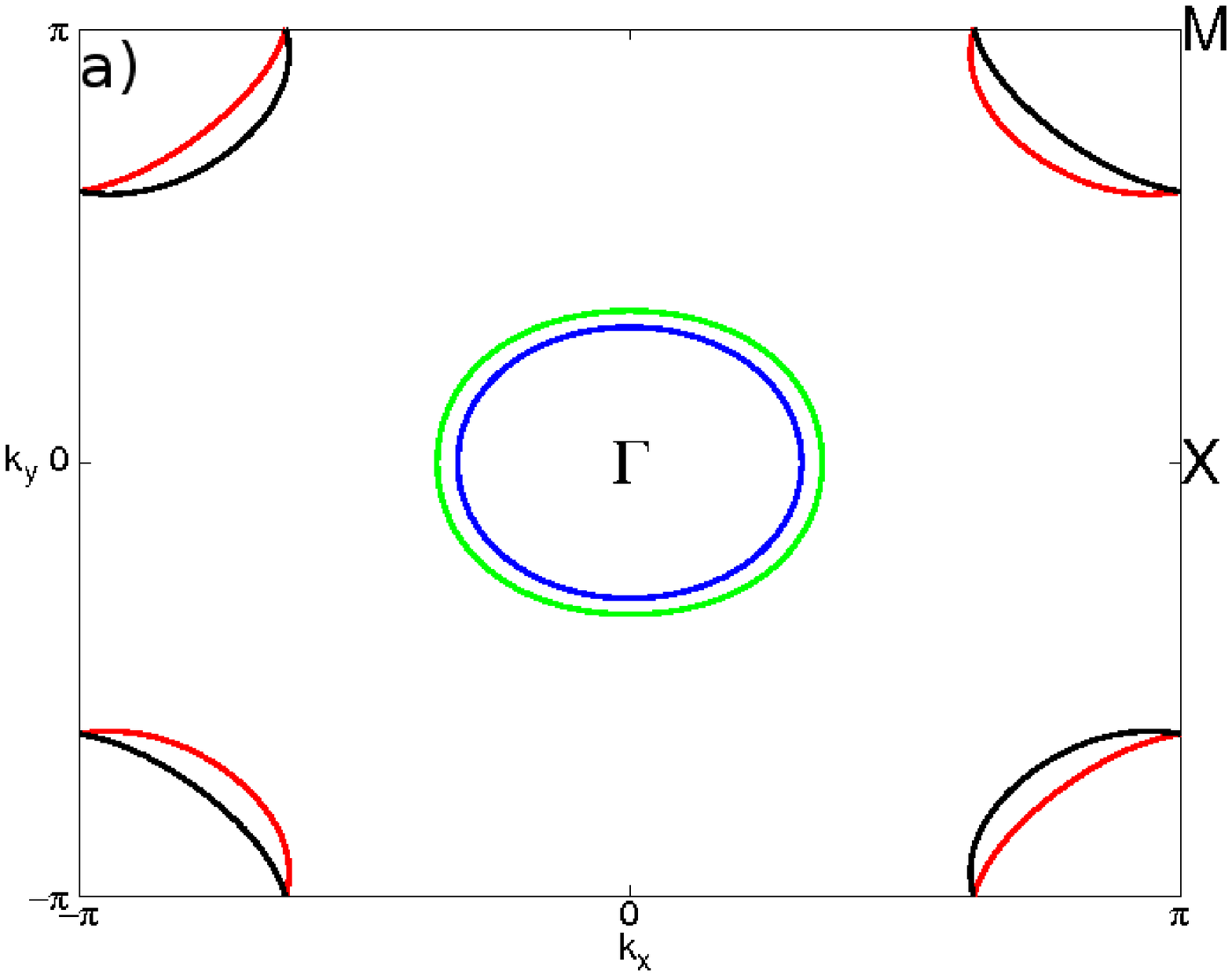}
\end{minipage}
\begin{minipage}[b]{1.5in}
\includegraphics[width=\textwidth]{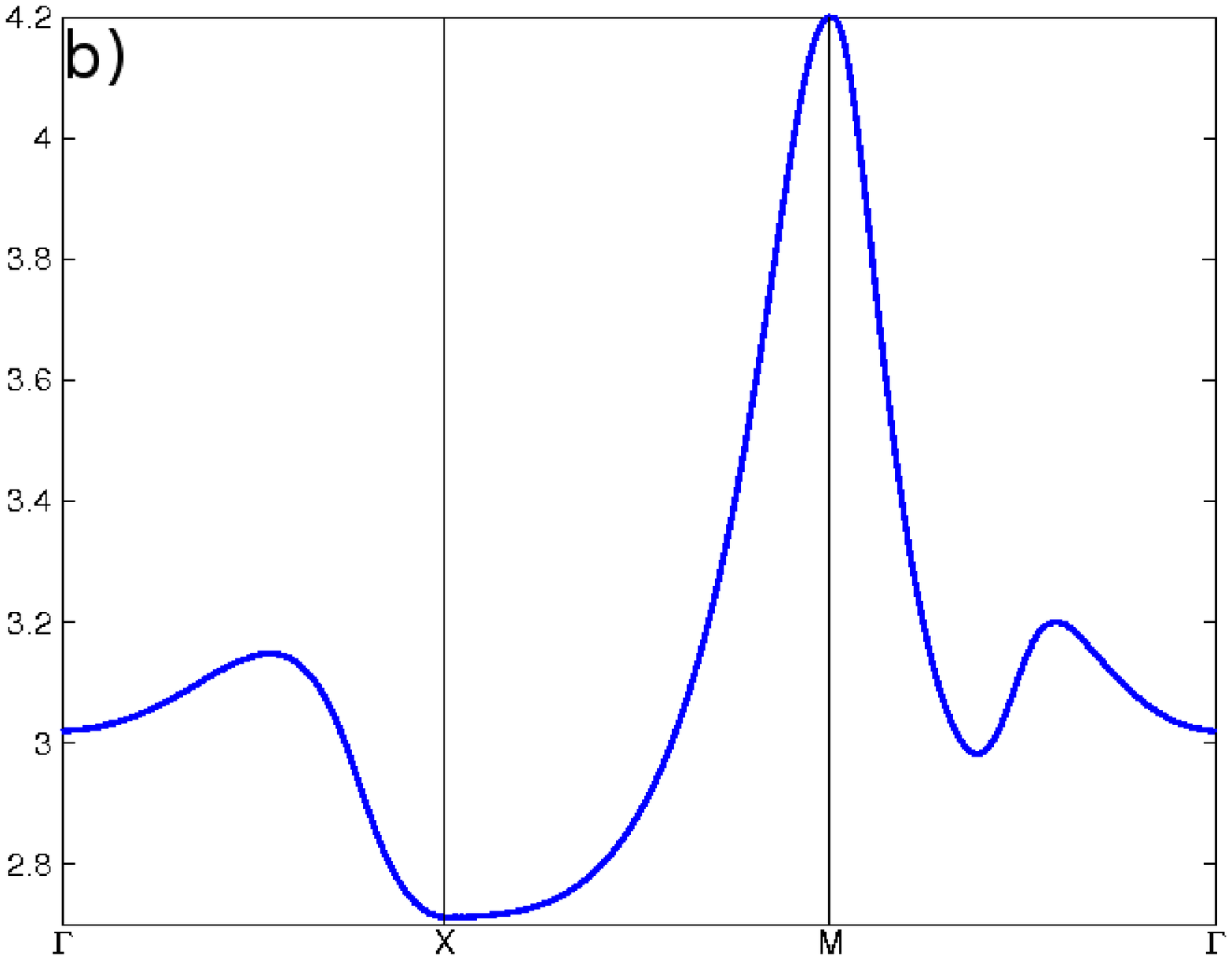}
\end{minipage}
\caption{(Color online) a) Fermi surface contour of the undoped pnictide compounds in our four-band model. b) Wavevector dependence of the static spin susceptibility along high symmetry lines. The peak indicates well nested electron-hole pockets at $\bm{Q}=(\pi,\pi)$.}\label{TB}
\end{figure}

\begin{figure}[t!]
\centering
\begin{minipage}[b]{1.5in}
\includegraphics[width=\textwidth]{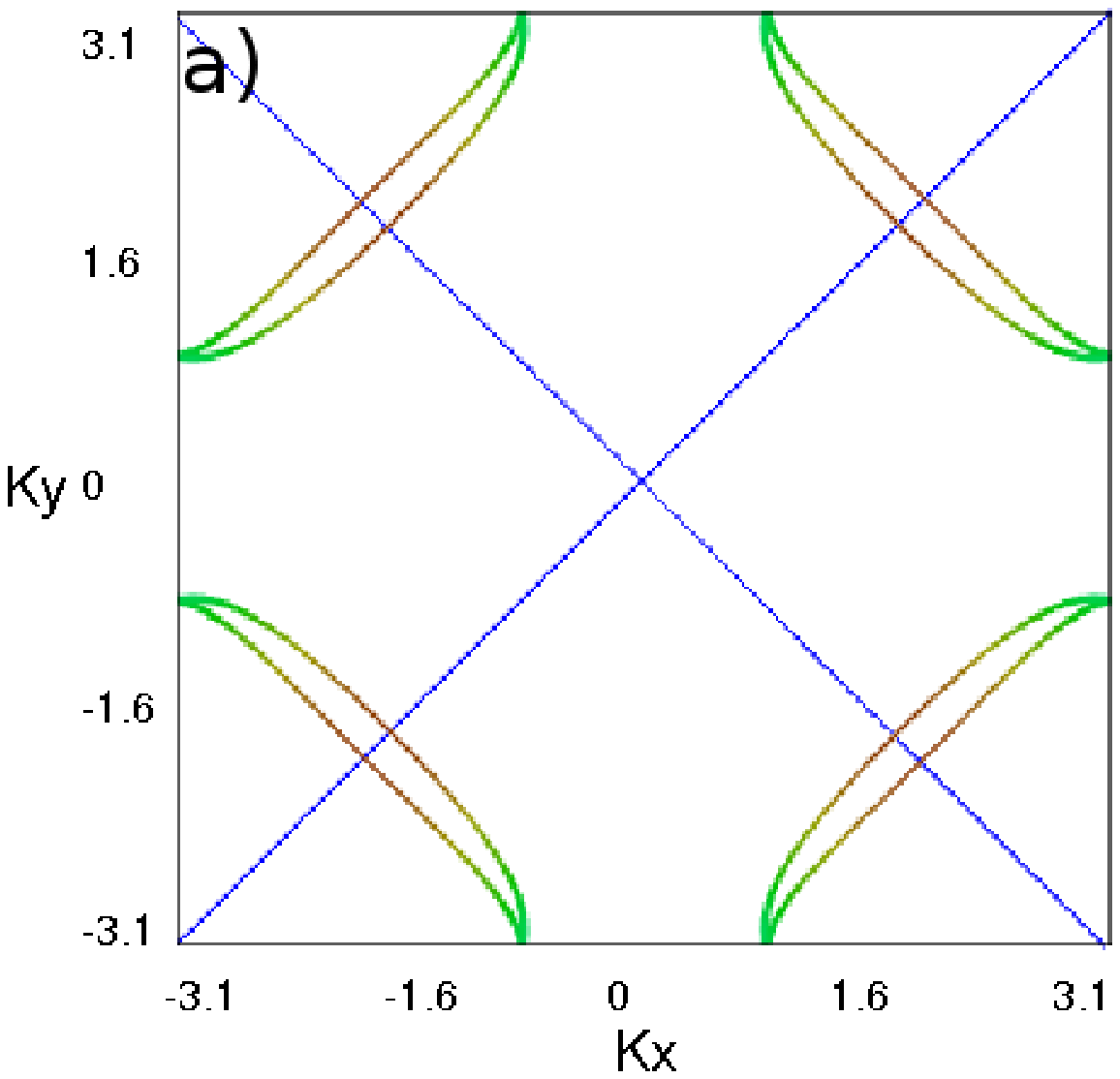}
\includegraphics[width=\textwidth]{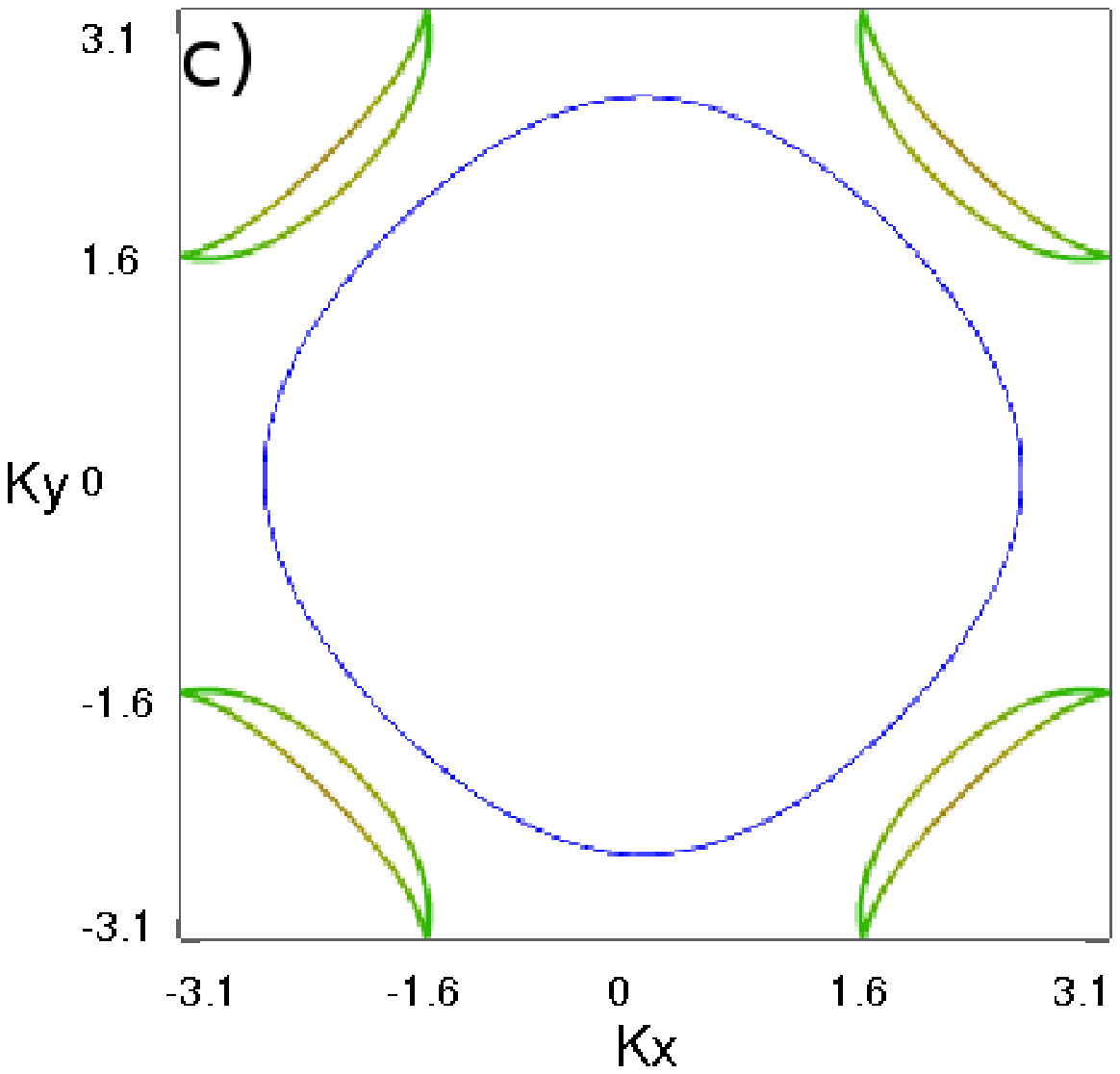}
\includegraphics[width=\textwidth]{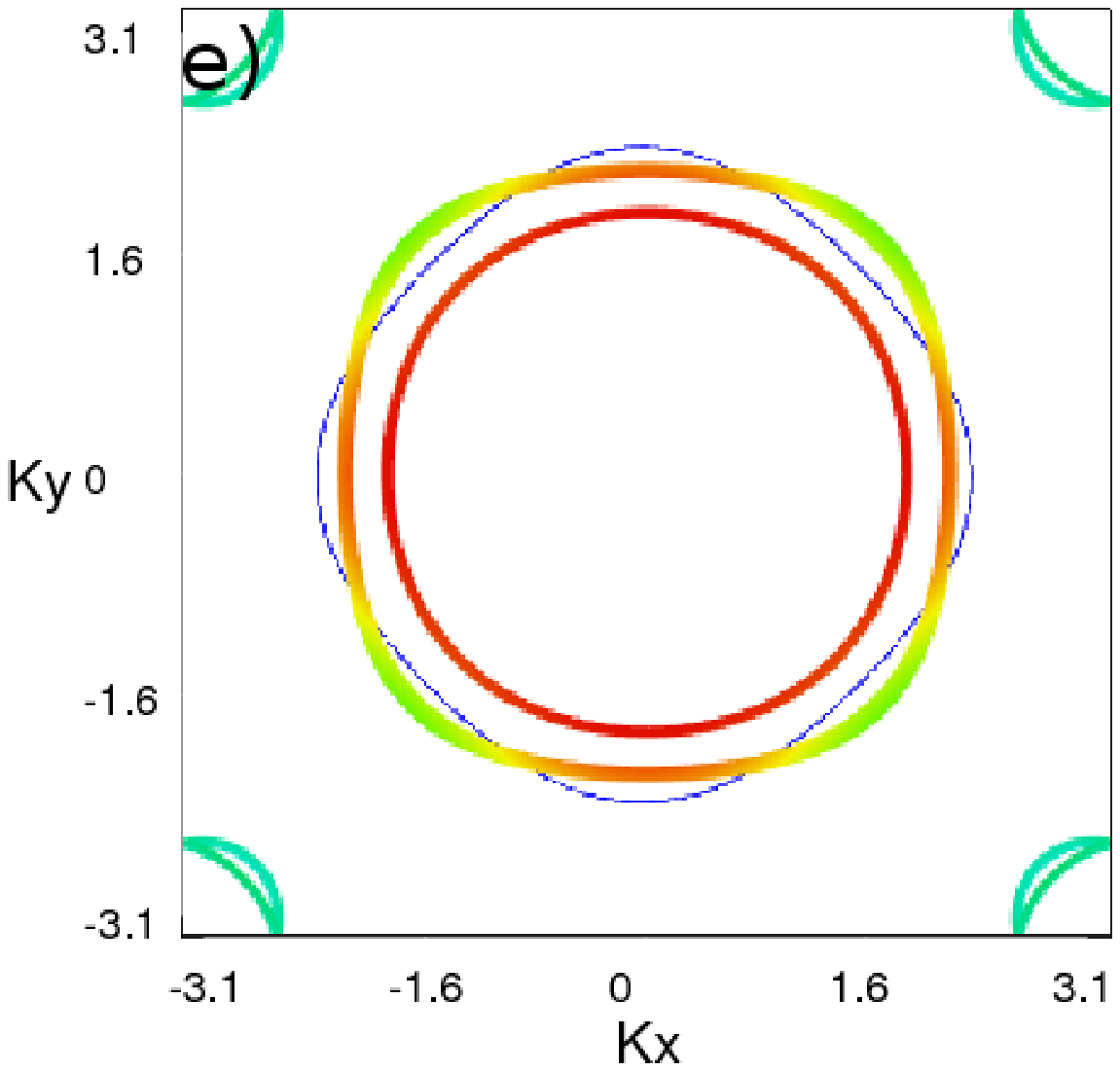}
\end{minipage}
\begin{minipage}[b]{1.5in}
\includegraphics[width=\textwidth]{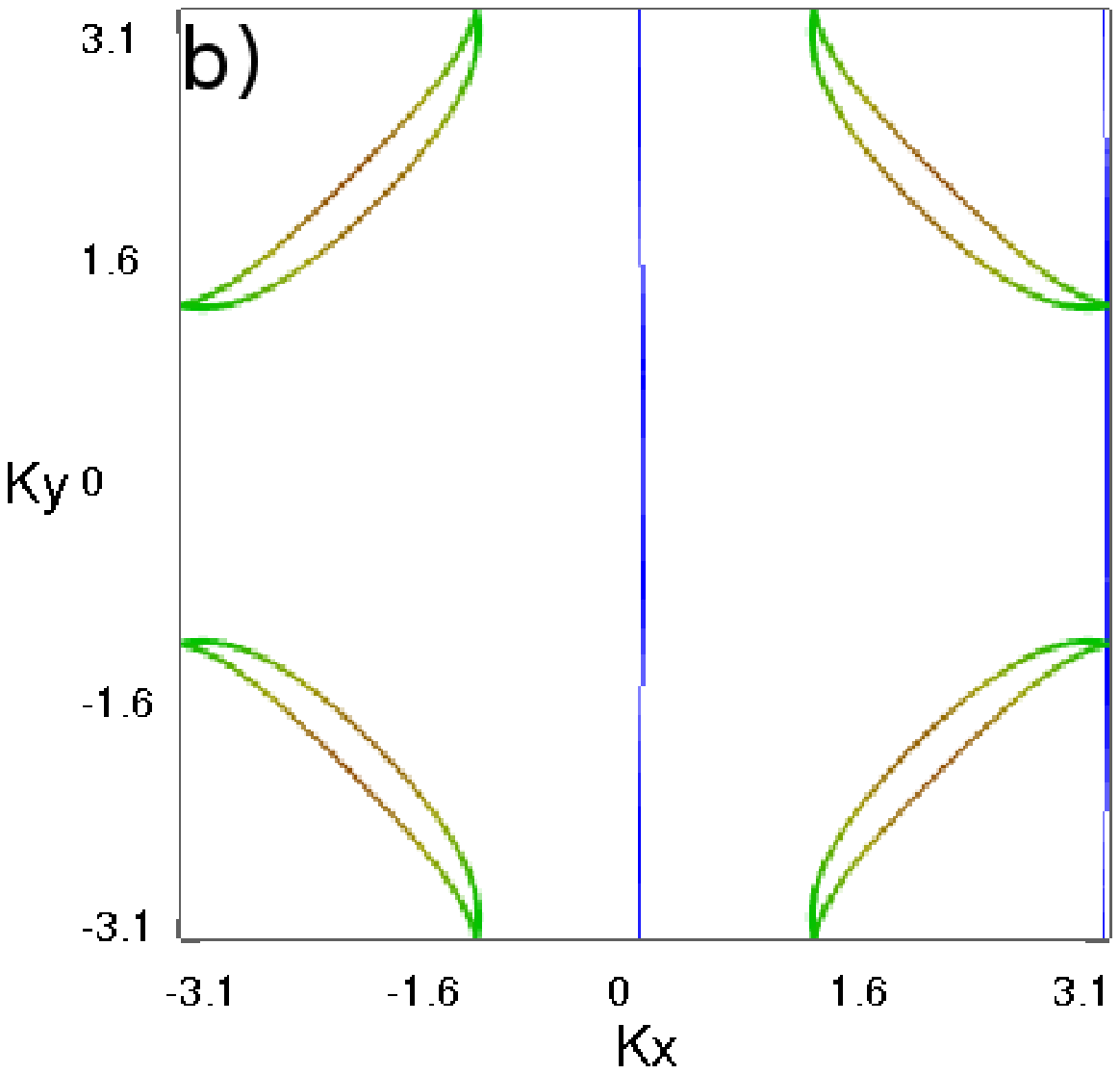}
\includegraphics[width=\textwidth]{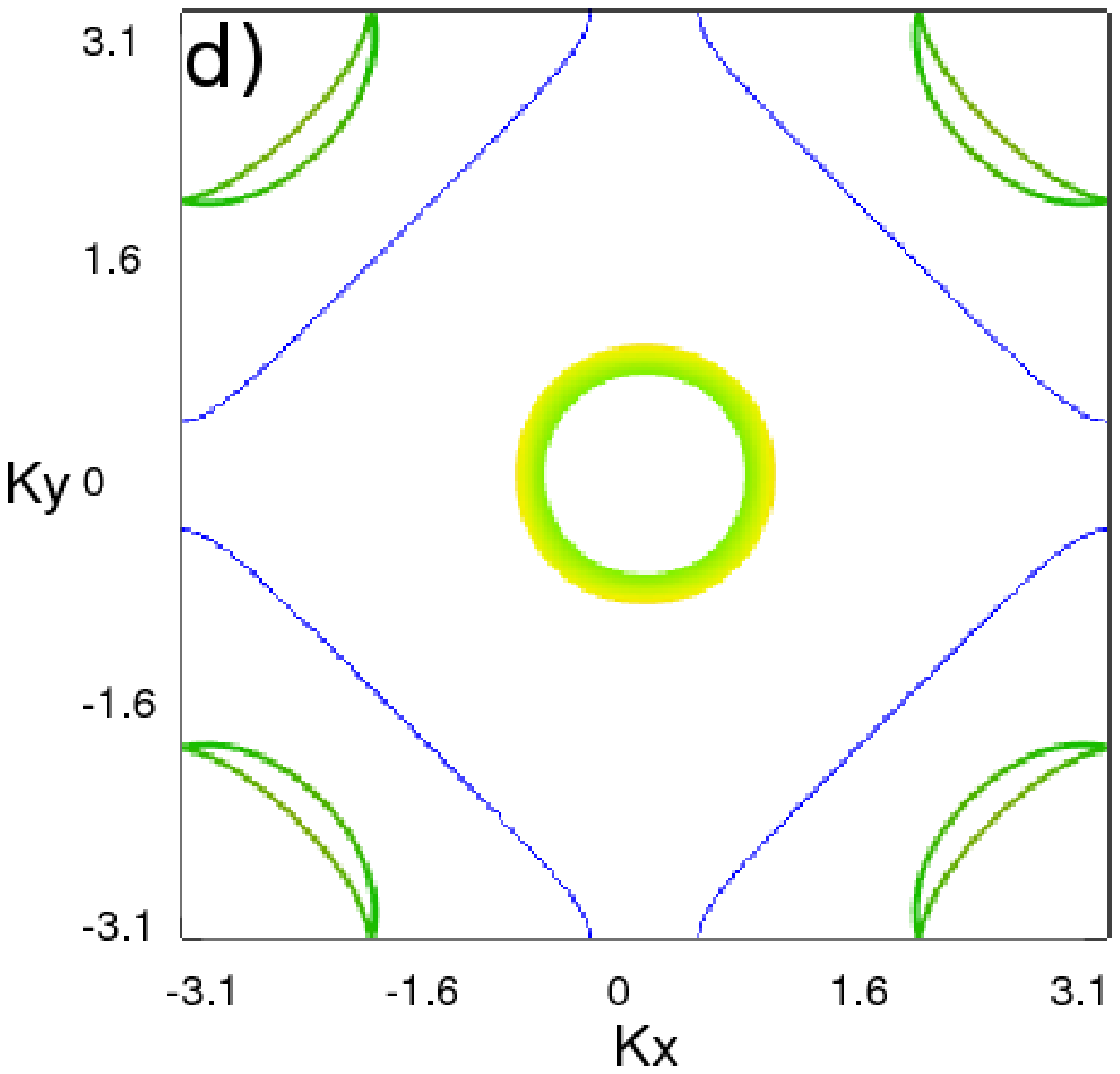}
\includegraphics[width=\textwidth]{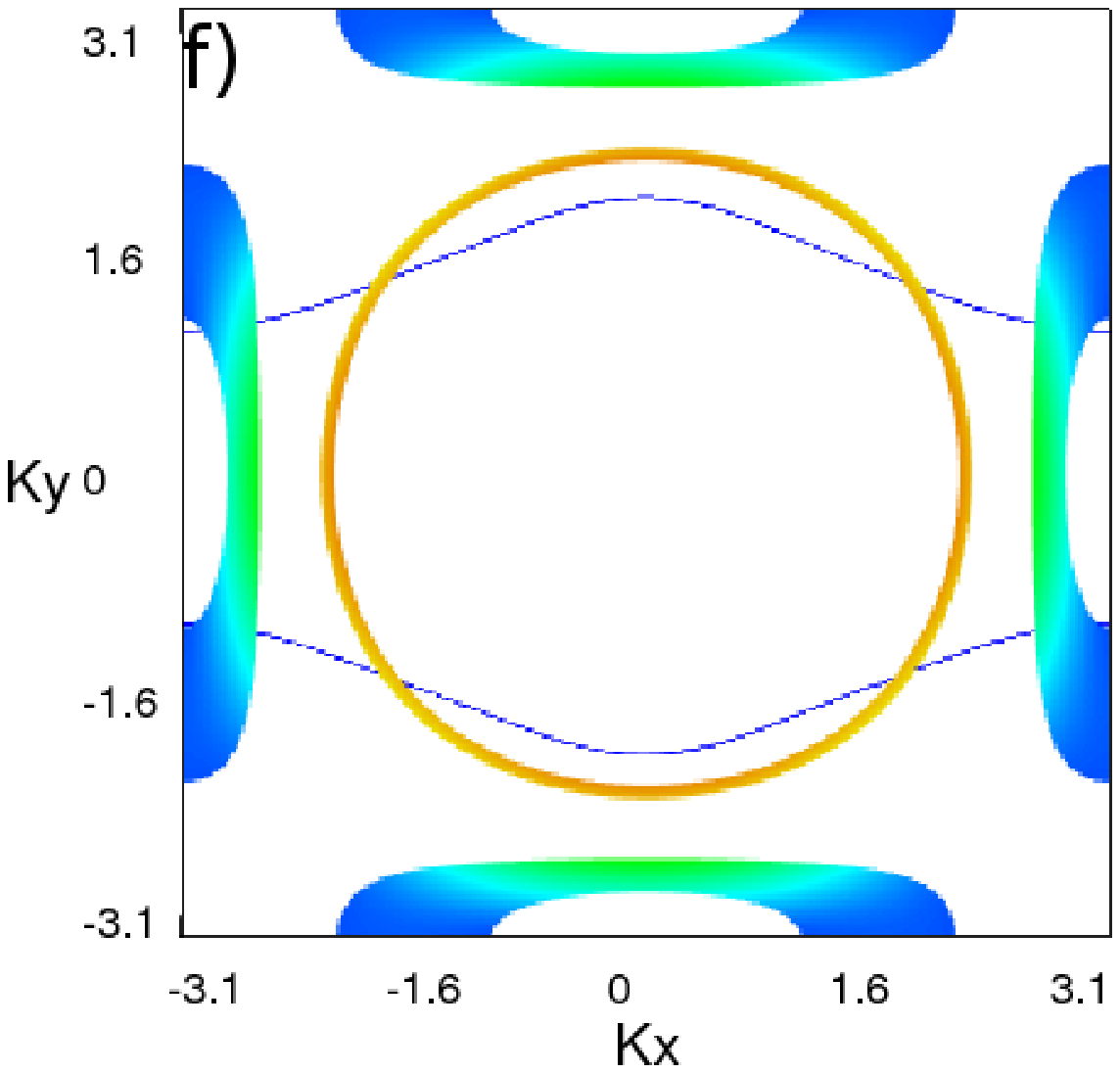}
\end{minipage}
\caption{(Color online) Contours of the FS, in the folded BZ, colored with arDOS$^{-1}$ 
 for $\mu=(1.0,0.75,0.45,0.1,-0.4,-0.6)$ from panel a) to panel f), respectively. Red FS lines means small arDOS. In the same panels we plot with blue lines the points where the respective, self-consistently obtained, SC gap vanishes ($q_c=\pi/6$). Around half filling we find an $s_{\pm}$ gap that persists in the electron ($\mu=0.1$ (d)) as well as hole doped region (see FIG. \ref{q3q4}c). The $s_{\pm}$ gap is {\it nodeless} with its vanishing area almost halfway between the electron and hole pockets and persists at further electron doping even when the hole bands do not contribute to the FS (c). More electron doping results in a cascade of transitions to {\it triplet} ($\sin k_x$) (b) and $d_{x^2-y^2}$-wave (a) solutions. For $\mu<0$, the nodes of the $s_{\pm}$ solution move closer to the outer hole pocket and finally intersect it (e). In the doping limit when the FS becomes like in (f) the SC gap is a superposition of $s_\pm+d_{x^2-y^2}$ symmetries. Note that the hole FS is thicker than the electron one due to its higher DOS, therefore  we have chosen to use a different color grade for the electron and hole bands.}\label{q3}
\end{figure}

Our aim is to describe the electronic structure of the iron pnictides around the Fermi energy as realistic as possible.
To this end we construct a four band tight-binding (TB) model which captures the essential physics of the undoped compounds, while combining features from the previously introduced TB models \cite{Eremin}. The energy bands are:
\begin{eqnarray}\nonumber
\epsilon_{1,2}^h &=& -t_1^{h_{1,2}}\left(\cos{k_x}+\cos{k_y}\right) -t_2^{h}\cos{k_x}\cos{k_y} -\epsilon_F^{h_{1,2}}\,,\nonumber\\
\epsilon_{1,2}^e &=& -t_1^e\left(\cos{k_x}+\cos{k_y}\right) -t_2^e\cos{k_x}\cos{k_y} -\epsilon_F^e\nonumber\\
&&\pm  t_3^e\sin{k_x}\sin{k_y}\,,
\end{eqnarray}
with $t_1^e=-1.14$, $t_2^e=-0.2$, $t_3^e=0.2$, $\epsilon_F^e=-1.5$, $t_1^{h_1}=-0.3$, $t_1^{h_2}=-0.2$, $t_2^{h}=-0.24$, $\epsilon_F^{h_1}=0.6$ and $\epsilon_F^{h_2}=0.4$. 
We have verified that these TB dispersions are in good agreement with band structure
calculations; not only the FS topology and the susceptibility are reproduced (see Fig.\ \ref{TB})
but also the angularly resolved density of states (arDOS) at any Fermi momentum $k_F$ is properly given (arDOS $\sim|\nabla\epsilon^{e,h}_{1,2}({\bf k}_F)|^{-1}$). To probe the doping dependence of the gap symmetry in our self-consistent calculations, we have introduced a chemical potential $\mu$ varying in the interval $-0.6\leq\mu\leq1.4$. When $\mu>0$ ($\mu<0$), the system is electron (hole) doped. The Fermi surfaces at various chemical potentials as well as the corresponding arDOS are depicted in Fig.\ \ref{q3}.

\par We solve the most general multiband Bardeen-Cooper-Schrieffer (BCS) equation at zero temperature \cite{Suhl},
\begin{equation}\label{nBCS}
\Delta_n({\bf k})=-\sum_{n',{\bf k'}}V_{nn'}({\bf k,k'})\frac{\Delta_{n'}({\bf k'})}{2E_{n'}({\bf k'})}\,,
\end{equation}
where $\Delta_n({\bf k})$ is the momentum dependent superconducting gap and $E_{n}({\bf k}) = \sqrt{\epsilon^2_{n}({\bf k})+\Delta_{n}^2({\bf k})}$ the quasiparticle excitation spectrum of the $n^{\rm th}$ band.
The pairing potentials $V_{nn'}({\bf k,k'})$ can support intra ($n=n'$) and interband ($n\neq n'$) scattering. As a first approximation, we will take $V_{nn'}({\bf k},{\bf k'})=V({\bf k}, {\bf k'})$. 
 This yields a "global" SC gap symmetry valid over the entire B.Z., which is nevertheless driven by the characteristics of each seperate band, such as the arDOS distribution. 

The effective pairing potential of the small-q phonon-mediated interaction takes the form \cite{smqGV2}: $V({\bf k},{\bf k'})=V_c^*-\frac{V_{ph}}{q^2_c+|{\bf k-k'}|^2}$ properly continued periodically, where a repulsive Coulomb pseudopotential $V_c^*$ is responsible for screening at short distances whereas the negative term is the attractive phonon part. The pairing kernel is characterized by a smooth momentum cutoff $q_c$ which selects the small wave vectors in the attractive phonon part while at larger wave vectors the repulsive Coulomb pseudopotential may prevail. Decreasing the cutoff $q_c$
leads to a situation that has been named {\it momentum decoupling} (MD)
\cite{smqGV1}. In the MD regime there is a tendency for superconductivity to
decorrelate in the various FS regions;
 the gap function gradually loses its rigidity in momentum space.
 In regions of the FS with high arDOS we observe higher gap amplitudes, whereas
 in the low arDOS regions the gap is smaller. This arDOS driven anisotropy of the gap
 is the fingerprint of MD \cite{smqGV1} and may be behind multiple gap signatures in the experiments.

To solve the BCS equation self-consistently with the momentum dependent pairing kernel, we perform the convolution integral in Eq.\ (\ref{nBCS}) by utilizing an FFT cyclic convolution technique.
The {\bf k}-summation was performed over a 512$\times$512 grid in a thin shell around the FS, in order to get maximal resolution. In the results presented here, the Coulomb pseudopotential is fixed at $V_c^*=0.09V_{ph}$ and the pairing amplitude at $V_{ph}=2$. Several values of the cutoff parameter $q_c$ were considered varying from $\pi$ to $\pi/8$ and various values of the chemical potential were explored. 

\begin{figure}[t!]
\includegraphics[width=0.295\textwidth]{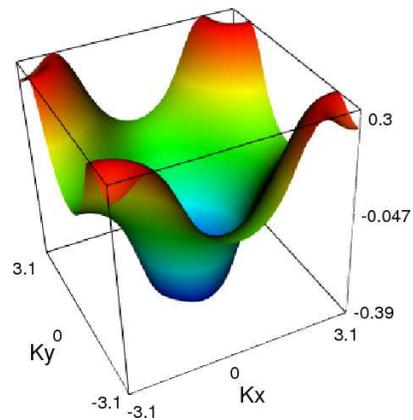}
\caption{(Color online) A typical, self-consistently obtained
 sign changing $s_{\pm}$ superconducting gap, originating from a small-q phonon interaction. The zeroes of this gap emerge between the electron and hole pockets (for $q_c=\pi/6$ and $\mu=-0.1$). Notice the anisotropy of the gap function and how the gap amplitude varies over different regions }\label{spm}
\end{figure}

When $q_c\gtrsim\pi/5$, anisotropic $s$-wave, sign-preserving solutions are found at all dopings. Further reducing $q_c$ yields interesting doping-induced transitions between different gap symmetries. 
When $\pi/5\geq q_c>\pi/8$ we get remarkable self-consistent solutions of the gap, depending on the doping level, that include nodal $s_\pm$+$d_{x^2-y^2}$-wave, $s_\pm$, $d_{x^2-y^2}$-wave and $p$-wave symmetries. Here we present calculations for $q_c=\pi/6$ as a representative example of our results.
Fig.\ \ref{q3} depicts the evolution of the FS, colored by arDOS$^{-1}$ as a function of doping. In the same plots, the lines where the respective self-consistently calculated SC gap vanishes are drawn.

\begin{figure}[t!]
\centering
\begin{minipage}[b]{1.1in}
\includegraphics[width=\textwidth]{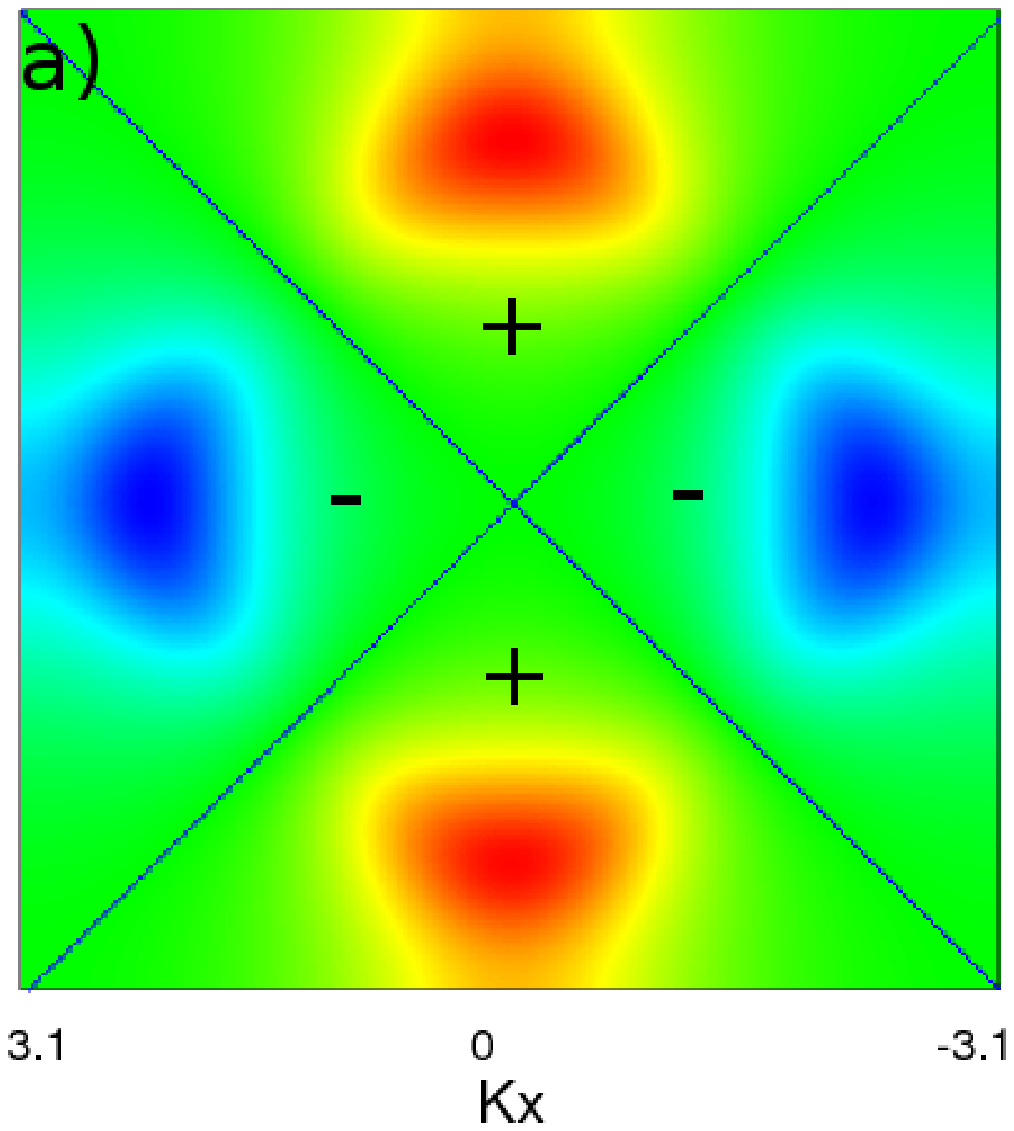}
\includegraphics[width=\textwidth]{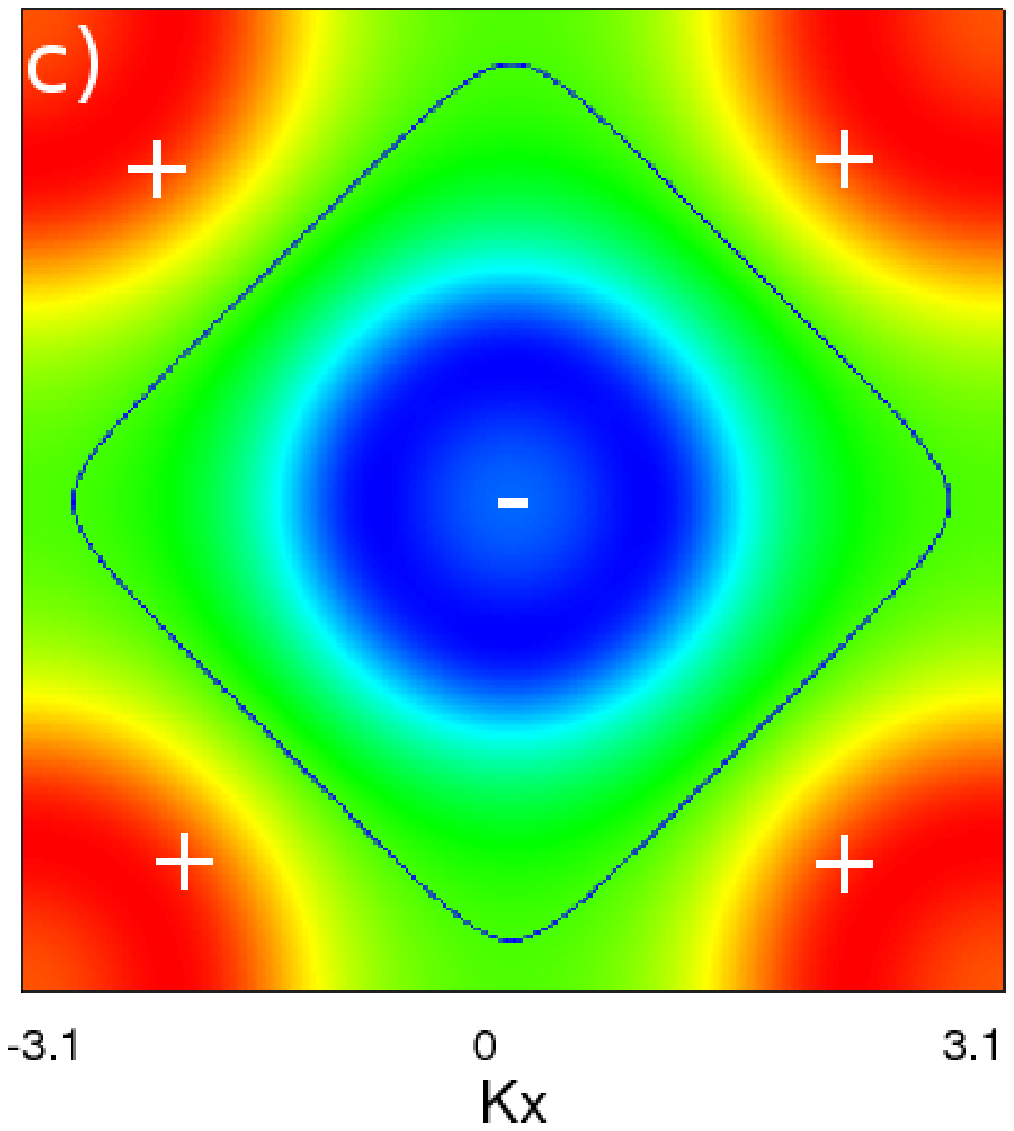}
\end{minipage}
\begin{minipage}[b]{1.1in}
\includegraphics[width=\textwidth]{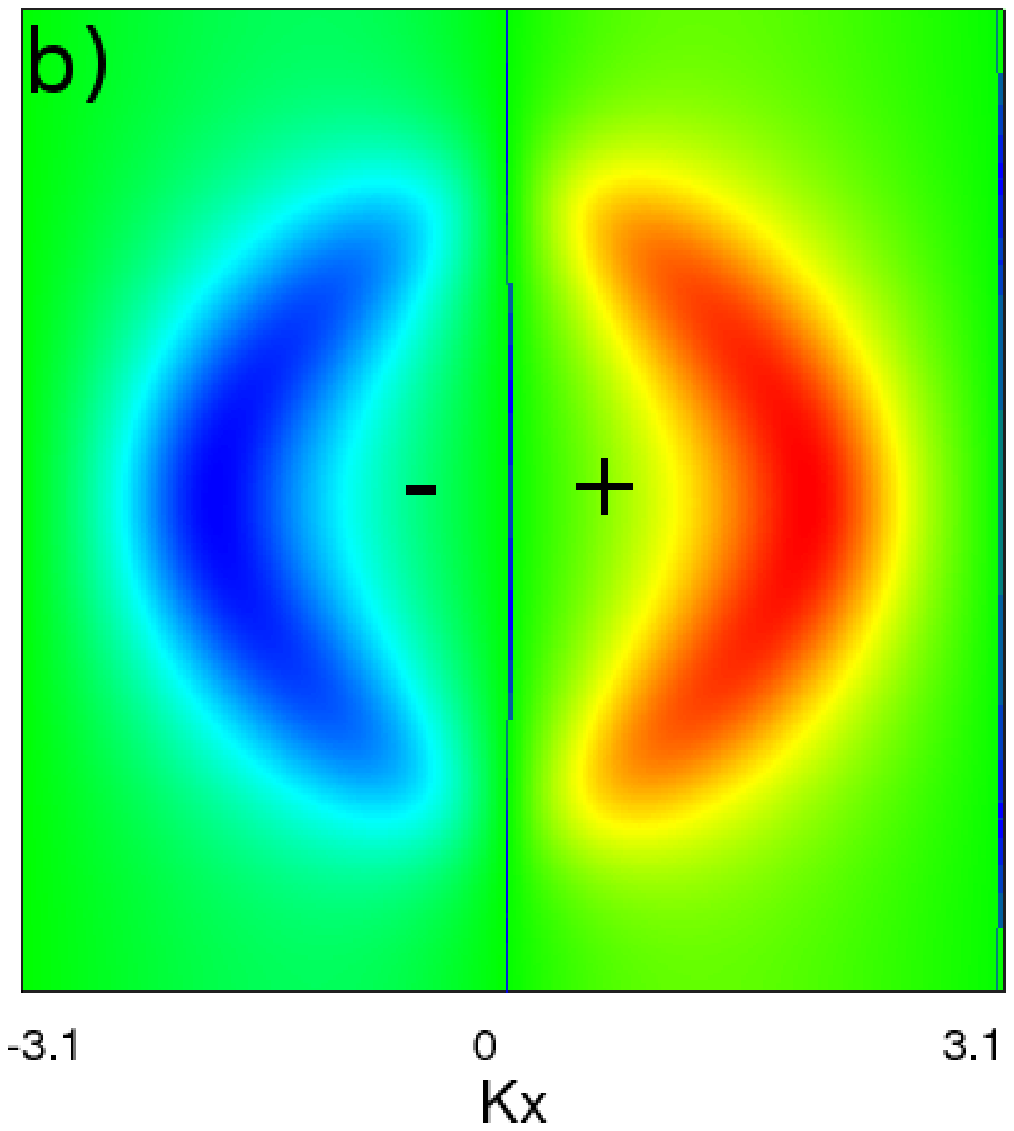}
\includegraphics[width=\textwidth]{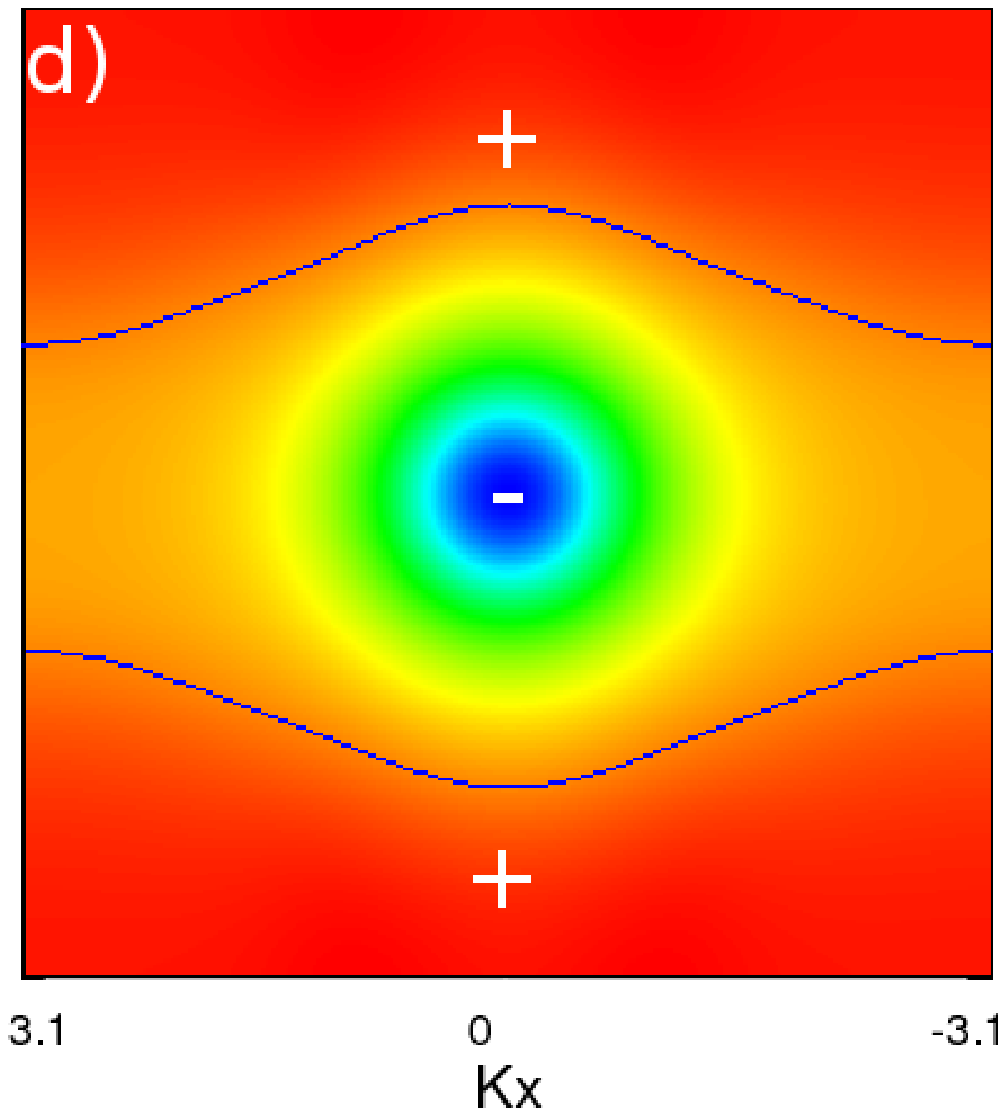}
\end{minipage}
\caption{(Color online) Color grade plots of characteristic superconducting gap amplitudes in the folded BZ
calculated self-consistently, when moving from the electron to the hole doped regime for  $q_c=\pi/6$. The panels a), b), c), and d) correspond to $\mu=1.15, 0.8, -0.1$, and $-0.6$, respectively. The corresponding FS shapes are depicted in Fig.\ \ref{q3}. Thin blue lines indicate the locations where the obtained SC gap vanishes. In the color grade, red color illustrates positive and blue a negative sign of the maximal gap.}\label{q3q4}
\end{figure}

Starting from the electron doped side, we find an anisotropic $d_{x^2-y^2}$-wave gap in the regime $1.4\geq\mu\geq 1.0$ as shown in Fig.\ \ref{q3q4}a. In this doping region, the high-DOS hole pockets lie away from the Fermi level, and only the electron pockets drive superconductivity. The SC gap contains nodes intersecting the FS in the lowest arDOS points (Fig.\ \ref{q3}a). The arDOS distribution at these dopings is highly anisotropic with the points where the gap nodes intersect the FS having considerably lowest DOS. Thus, gapping the rest of the FS while having nodes at the lowest DOS points is energetically favorable for SC.

Decreasing $\mu$ shrinks the electron pockets and makes the arDOS more isotropically distributed. When $0.95\geq\mu\geq 0.65$, a structure of the $p$-wave ($\sin{k_x}$) type develops with nodes at the locations where the electron pockets cross (Figs.\ \ref{q3} and \ref{q3q4}b). Although these FS points have maximal arDOS this is only approximately two times larger than the minimum arDOS on the FS. The system can still minimize its energy by keeping the rest of the FS gapped while leaving nodes on these points. Since superconductivity in our approach originates from particle pairing in the same band, the symmetry of the gap function implies a {\it spin triplet} order parameter. Such a state has already been proposed for the iron pnictides in a different context \cite{P.A.Lee}. However, a spin-singlet $p$-wave order parameter is also possible if one considers interband pairing \cite{F.C.Zhang}.

In the interval $0.6\geq\mu\geq 0.25$ a nodeless $s_{\pm}$ gap is found, yet with the nodal lines close to the electron pockets (Fig.\ \ref{q3}c). The anisotropy of the arDOS is further reduced and any node on the FS is not the preferred configuration. This $s_\pm$ solution is mainly a superposition of $\cos{k_x}\cos{k_y}+(\cos{k_x}+\cos{k_y})$ representations with different weights. As we move towards half filling, the hole pockets approach the Fermi level, the $\cos{k_x}\cos{k_y}$ component is suppressed and the nodes of the gap move away from the electron pockets.

Around half filling, $0.2\geq\mu\geq-0.3$, the calculations converge to a SC gap of the form shown in Fig.\ \ref{q3}d which is of the \textit{nodeless} $s_{\pm}$ type, cf.\ Fig.\ref{q3q4}c. In this regime, the FS consists of both electron and hole bands and the node lines lie almost halfway between them. This happens due to MD phenomena and the presence of interband coupling between the bands. The latter makes the SC gaps over the hole and electron FS coupled while the former ensures that maximum energy is gained by leaving the empty (no FS) space between these FS sheets ungapped. Such a mechanism results in a sign alternating $s_\pm$ solution over the entire B.Z. The gap is nearly isotropic or may exhibit some arDOS driven momentum anisotropy (see Fig.\ref{spm}) and corresponds to the nearly isotropic BCS-like gap reported in several experiments \cite{BCS-likeSmFeAsF,Ding,Kondo}. 

Upon further doping the system with holes, $-0.35\geq\mu\geq-0.55$, the nodes of the $s_\pm$ gap intersect the outer hole FS (Fig.\ref{q3}e). 
Finally, at $\mu=-0.6$ the FS and the arDOS distribution become as shown in Fig.\ \ref{q3}f. The gap now is a superposition of $s_\pm$ and $d_{x^2-y^2}$ symmetries
with nodes at the low arDOS points of the FS. This rich structure agrees with previous theoretical studies suggesting the near degeneracy of these two symmetries \cite{Graser}. Along with the pure $d_{x^2-y^2}$-wave  $p$-wave and $s_\pm$ solutions, this gap structure offers a possible explanation to NMR and penetration depth reports of nodal superconductivity in both electron and hole doped compounds \cite{Martin,Grafe}. Notice that our theory predicts transitions from nodeless to nodal SC when going from the underdoped to the overdoped regime, (either with hole or electron doping). Exactly such a behavior has recently been found experimentally \cite{YukeLi}.

In conclusion, we have self-consistently computed the BCS gap of electron and hole doped iron pnictides.  Our study offers a scenario for understanding the gap symmetry of the superconducting state in these materials. We have shown that small-q electron-phonon scattering can accommodate conflicting experimental results producing a gap that changes sign while being compatible with the observed isotope effect. When both electron and hole FS sheets are present, we obtain within this purely phononic mechanism as self-consistent solution the fully gapped $s_{\pm}$ state, which was so far considered as evidence for a spin-fluctuations mechanism.
Depending on doping, the computed gap symmetries vary from nodeless to nodal in agreement with recent observations. Our solutions include $s_{\pm}+d$-wave in the hole doped side and $d$-wave, as well as spin-triplet gap functions in the electron doped side.
The present results establish that the small-q phonon picture for
the pairing mechanism in the iron pnictides can explain the conflicting observations and, hence, is the plausible one.
Understanding the competition of SC with spin density
waves as well as magnetoelastic effects when forward processes dominate the electron-phonon
interaction could provide further insight.

\begin{acknowledgments}
We are grateful to J.G. Analytis for enlightening discussions. A.A. and P.K. acknowledge financial support by the N.T.U. Athens, P.K from the EU project NanoCTM and P.M.O. from the Swedish Research Council (VR).
\end{acknowledgments}


\begin{thebibliography}{999}
\bibitem{Kamihara} Y. Kamihara {\it et al.}, J. Am. Chem. Soc. \textbf{130}, 3296 (2008).
\bibitem{Mazin2} I.I. Mazin and J. Schmalian, Physica C \textbf{469}, 614 (2009) and references therein.
\bibitem{NatureNeutrons} A.D. Christianson {\it et al.}, Nature {\bf 456}, 930 (2008).

\bibitem{IsotopeNat} R.H. Liu {\it et al.}, Nature \textbf{459}, 64 (2009).

\bibitem{BCS-likeSmFeAsF} T.Y. Chen {\it et al.}, Nature \textbf{453}, 1224 (2008).
\bibitem{Ding} H. Ding {\it et al.}, EPL {\bf 83}, 47001 (2008).
\bibitem{Kondo} T. Kondo {\it et al.}, Phys. Rev. Lett. {\bf 101}, 147003 (2008).

\bibitem{Martin} C. Martin {\it et al.}, Phys. Rev. Lett. {\bf 102}, 247002 (2009).
\bibitem{Malone} L. Malone {\it et al.}, Phys. Rev. B {\bf 79}, 140501(R) (2009).
\bibitem{Grafe} H-J. Grafe {\it et al.}, Phys. Rev. Lett. {\bf 101}, 047003 (2008).
\bibitem{Yashima} M. Yashima {\it et al.}, J. Phys. Soc. Jpn. {\bf 78}, 103702 (2009).
\bibitem{YukeLi} Y. Li {\it et al.}, New J. Phys. {\bf 12}, 083008 (2010).
\bibitem{Singh1} D.J. Singh and M.H. Du, Phys. Rev. Lett. \textbf{100}, 237003 (2008); K. Haule, J.H. Shim and G. Kotliar, Phys. Rev. Lett. \textbf{100}, 226402 (2008); Z.P. Yin {\it et al.}, Phys. Rev. Lett. \textbf{101}, 047001 (2008); D.J. Singh, Phys. Rev. B \textbf{78}, 094511 (2008); C. Cvetkovic and Z. Tesanovic, EPL \textbf{85}, 37002 (2009).
\bibitem{Mazin1} I.I. Mazin {\it et al.}, Phys. Rev. Lett. \textbf{101}, 057003 (2008).

\bibitem{Clarina} Clarina de la Cruz {\it et al.}, Nature \textbf{453}, 899 (2008). 
\bibitem{Kuroki} K. Kuroki {\it et al.}, Phys. Rev. Lett. \textbf{101}, 087004 (2008); Z.J Yao, J.X. Li and Z.D. Wang, New J. Phys. \textbf{11}, 025009 (2009); K. Seo, B.A. Bernevig, and J. Hu, Phys. Rev. Lett. \textbf{101}, 206404 (2008);  F. Wang {\it et al.}, Phys. Rev. Lett. \textbf{102}, 047005 (2009);  R. Sknepnek {\it et al.}, Phys. Rev B \textbf{79}, 054511 (2009).


\bibitem{Graser} S. Graser {\it et al.}, New J. Phys. \textbf{11}, 025016 (2009); R. Thomale, C. Platt, J. Hu, C. Honerkamp and B. A. Bernevig, Phys. Rev. B \textbf{80}, 180505(R) (2009).
\bibitem{Nakai} Y. Nakai {\it et al.}, Phys. Rev. B \textbf{79}, 212506 (2009).
\bibitem{Granath} M. Granath {\it et al.}, Phys. Rev. B \textbf{79}, 235103 (2009)
\bibitem{smq1} K. J. von Szczepanski and K. W. Becker, Z. Phys. B \textbf{89}, 327 (1992); M. L. Kulic and R. Zeyher, Phys. Rev. B \textbf{49}, 4395 (1994); 
  A. A. Abrikosov, Physica \textbf{222C}, 191 (1994).
\bibitem{smqGV1} G. Varelogiannis, Phys. Rev. B \textbf{57}, 13 743 (1998).
\bibitem{Abrikosov} A. A. Abrikosov, Phys. Rev. B \textbf{53}, R8910 (1996); \textbf{56}, 446 (1997); A. J. Leggett, Phys. Rev. Lett. \textbf{83}, 392 (1999); M. Weger and M. Peter, Physica (Amsterdam) \textbf{317C-318C}, 252 (1999).
\bibitem{smqGV2} G. Varelogiannis \textit{et al.}, Phys. Rev. B \textbf{54}, R6877 (1996); G. Varelogiannis, Phys. Rev. B \textbf{57}, R732 (1998); A. Perali and G. Varelogiannis, Phys. Rev. B \textbf{61}, 3672 (2000).

\bibitem{Agterberg} D. F. Agterberg, V. Barzykin, and L. P. Gorkov, Phys. Rev. B \textbf{60}, 14 868 (1999); P.M. Oppeneer and G. Varelogiannis, Phys. Rev. B \textbf{68}, 214512 (2003).
\bibitem{smqGV3} G. Varelogiannis, Phys. Rev. Lett. \textbf{88}, 117005 (2002).
\bibitem{Cobaltite} Xiao-Shan Ye, Zi-Jian Yao, and Jian-Xin Li, J. Phys.: Condens. Matter {\bf 20}, 045227 (2008).
\bibitem{Eremin} M.M. Korshunov and I. Eremin, Phys. Rev. B \textbf{78}, 140509(R) (2008); S. Raghu \textit{et al.}, Phys. Rev. B \textbf{77}, 220503(R) (2008).

\bibitem{Suhl} H. Suhl, B.T. Matthias and L.R. Walker, Phys. Rev. Lett. \textbf{3}, 552 (1959).
\bibitem{P.A.Lee} P.A. Lee and X-G. Wen, Phys. Rev. B \textbf{78}, 144517 (2008).
\bibitem{F.C.Zhang} X. Dai {\it et al.}, Phys. Rev. Lett. \textbf{101}, 057008 (2008).
\end{thebibliography}
\end{document}